\documentclass[12pt]{iopart}
\usepackage{graphicx}
\usepackage{iopams}
\begin{document}

\title{Reconfigurable controlled two-qubit operation on a quantum photonic chip}

\author{H W Li$^1$, S Przeslak$^1$, A O Niskanen$^1$, J C F Matthews$^2$, A Politi$^2$, P Shadbolt$^2$, A Laing$^2$, M Lobino$^2$, M G Thompson$^2$ and J L O'Brien$^2$}
\address{$^1$Nokia Research Center, Broers Building, 21 JJ Thomson Avenue, Cambridge CB3 0FA, UK}
\ead{antti.niskanen@nokia.com}
\address{$^2$Centre for Quantum Photonics, H. H. Wills Physics Laboratory \& Department of Electrical and Electronic Engineering,
University of Bristol, Merchant Venturers Building, Woodland Road, Bristol, BS8 1UB, UK}
\ead{jeremy.obrien@bristol.ac.uk}

\begin{abstract}
Integrated quantum photonics is an appealing platform for quantum information processing, quantum communication and quantum metrology.  In all these applications it is necessary not only to be able to create and detect Fock states of light but also to program the photonic circuits  that implements some desired logical operation. Here we demonstrate a reconfigurable controlled two-qubit operation on a chip using a multiwaveguide interferometer with a tunable phase shifter. We find excellent agreement between theory and experiment, with a 0.98 $\pm$ 0.02 average similarity between measured and ideal operations.
\end{abstract}

\maketitle

Experimental quantum information processing is being pursued following several different paradigms and using numerous different physical realisations \cite{ladd, johnson}.
The ability to miniaturize and implement complicated optical experiments on an inherently stable and programmable chip makes integrated quantum photonics an attractive and feasible technology. Linear quantum photonics is ideally suited for quantum communication \cite{qkd} and interferometric quantum metrology \cite{metro}. In particular, in order to build quantum information processing devices following the qubit paradigm in linear optics, the best known way to implement entangling gates is using a probabilistic approach \cite{klm,ralph}. While in other physical realisations universal two-qubit gates such as the iSWAP \cite{martinis,niskanen} or perhaps even the B-gate\cite{whaley} might be more easily accessible, in the linear optics framework the controlled-NOT (CNOT) operation is achieved very naturally \cite{klm}. 

\begin{figure*}[ht]
\begin{center}
\includegraphics[width=1\textwidth]{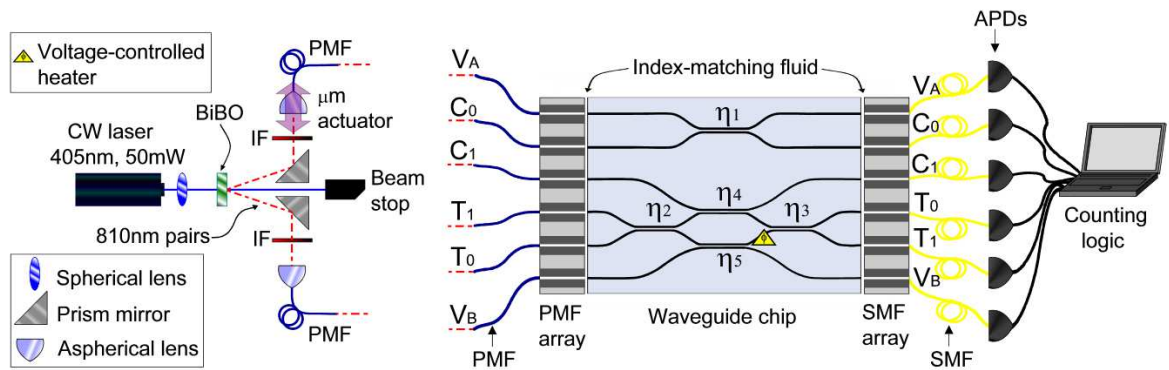}
\caption{\label{fig1}Setup for a reconfigurable controlled quantum gate. The 2.4 cm long chip is made using silica on silicon. The 3.5 $\mu$m by 3.5 $\mu$m single mode waveguides were buried 16 $\mu$m under the surface of the chip. The waveguides are separeted by 250 $\mu$m from each other at the input and output facets. The input and output waveguide arrays with 250 $\mu$m pitch were carefully aligned using an alignment stage. The coupling was further improved using index matching fluid. The voltage controlled phase shifter is a resistive thin-film heater on top of the waveguide. It is connected to electronics by standard wire bonding. The design reflectivities are $\eta_1=\eta_4=\eta_5=1/3$ and $\eta_2=\eta_3=1/2$. The reflectivities were measured as $\eta_1=0.324\pm 0.008$, $\eta_2=0.435\pm 0.015$, $\eta_3=0.469\pm 0.009$, $\eta_4=0.317\pm 0.007$ and $\eta_5=0.298\pm 0.012$.} 
\end{center}
\end{figure*}

Here we focus on the implementation of reconfigurable photonic quantum gates on chip \cite{MatthewsNP,shor,smith}. We study a tunable two-qubit gate that is a generalisation of the controlled-NOT gate. We use the path-encoding of qubits as opposed to polarisation encoding \cite{pol} which is another emerging approach to integrated quantum photonics. We demonstrate that one can readily generate a whole range of entangling two-qubit operations on such programmable chips. While in standard textbook implementations of quantum algorithms typically just the CNOT is used, such reconfigurable entangling gates are natural building blocks for complicated quantum algorithms.  Having access to a wide range of programmable two-qubit gates makes the compilation of quantum algorithms easier. It is always desirable to make quantum circuits as compact as possible to avoid decoherence that may effect computation. Moreover, this type of two-qubit devices will find immediate applications in quantum cryptography and quantum measurements, especially in miniaturized environments. Our results are a step forward in optics-based quantum information processing \cite{nphys,ncomm}.

The on-chip device implementing the tunable two-qubit operation was designed to be an extension of the CNOT gate \cite{politi}. It has six input and output waveguides, two of which are auxiliary at both ends. The waveguides denoted $C_0$, $C_1$, $T_0$ and $T_1$ encode the states 0 and 1 for the control and target qubits, respectively. The intended total photon number in the experiment is two: we only pay attention to two-photon coincidences. For instance, the presence of a photon in waveguides $C_0$ and $T_1$ each means that the qubit configuration is $|01\rangle$. Not all of the possible two-photon Fock states are logically meaningful: There should be precisely one photon in the control qubit waveguides and one photon in the target qubit waveguides. Appropriate configurations can be post-selected based on coincidence counting. 

Figure~\ref{fig1} illustrates the experimental setup. The reconfigurable quantum photonic chip was coupled from both sides to V-groove arrays of optical fibres. On the input side the chip was butt-coupled with index matching fluid to polarization maintaining fibres, while on the output side single mode fibres were used. The coupling efficiency through the chip (including input and output loss) varied between 50-65\% in the different experiments reported here. The type-I spontaneous parametric down-conversion source was pumped using a 50 mW 405 nm laser.
The 810 nm vertically polarized daughter photons were collected from the BiB$_3$O$_6$ crystal into two polarization maintaining fibres (see Fig 1) through 2nm bandpass filters and aspheric lenses. The pair of PM fibres could then be coupled to any pair of the waveguides in the waveguide arrays. After the photons passed through the chip they could be detected using fibre-coupled silicon avalanche photodiodes (APDs) which have greater than 50\% efficiency. The resulting voltage pulses were sent to an FPGA-based counting card capable of time tagging up to 16 independent channels. Since all the coincidence counting was done in software, it was possible to count coincidences simultaneously between all pairs of output fibres. When the photon pair source was connected directly to the APDs, about 11000 coincidences per second were typically observed. In this case each APD recorded about 80000 photons per second. The probability of multipair generation for the relevant time window (typically 3-5 ns, adjustable in software) is below 1\%.

The quantum photonic chip was fabricated using silica-on-silicon technology \cite{politi}. The on-chip voltage-controlled thermal phase shifter allowed for programming the chip to implement a continuous range of controlled operations \cite{MatthewsNP}. In practise the phase-controlled experiments were performed in 1~s pulsed intervals with 1-10~s cooling periods. Pulsing the voltage-controlled phase shifter while measuring either bright laser light power or photon pair coincidences enabled the determination of the phase voltage relation. The phase could be thereby set to any desired value. Voltages in the range 0-7 V were sufficient to reach a full $2\pi$ modulation. The practical reason for using voltage pulses was better long-term thermal stability compared to exploiting dc voltages to program the phase shifter. Pulsing effectively leads to a reduced counting duty cycle while the laser is on continuously. We only pay attention to what happens during the pulse. 

To establish the quantum nature of the experiment we performed a range of control experiments. We could readily prove the indistinguishability of the photons by carrying out
a Hong-Ou-Mandel \cite{hom} (HOM) coincidence measurement using the bottom coupler between waveguides $C_0$ and $V_A$. The reflectivity of this coupler is nominally 1/3. We delayed the arrival of one of the photons by finely scanning the position of one of the fibre launches. When the optical path lengths precisely matched, we observed a pronounced dip with visibility of $73.1\% \pm 1.6\% $ in the photon coincidence rate. The theoretical expectation of the dip visibility $V$ is $71.9\pm 2.9\%$ based on the independently measured reflectivity of $\eta_5=0.298\pm 0.012$ and 
\begin{equation}\label{vis}
V=1-(2\eta-1)^2/(\eta^2+(\eta-1)^2).
\end{equation}
This expression is obtained by considering the coincidence rate between distinguishable photons compared to indistinguishable photons.
The experimental result indicates that the quality of the indistinguishable photon source is high. 

\begin{figure}[h]
\begin{center}
\includegraphics[width=1\textwidth]{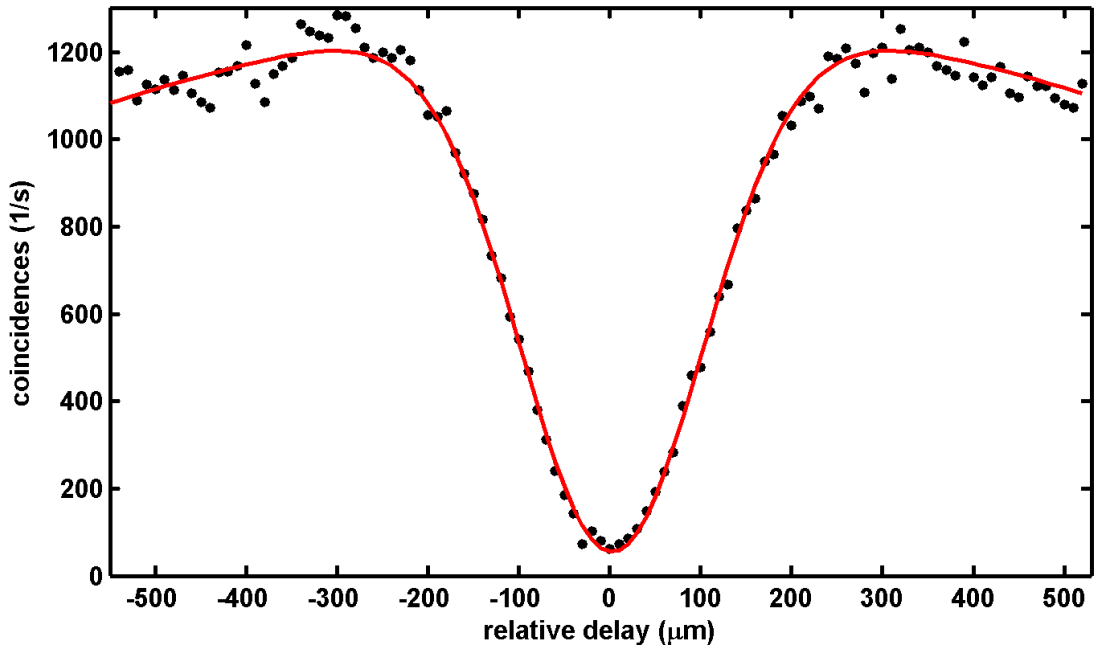}
\caption{\label{hom} Hong-Ou-Mandel effect in the second nominally 1/2 reflectivity directional coupler. The solid line is a Gaussian fit with a sinc component yielding a visibility of $95.8 \% \pm 1.6 \%$. The dip width ($\sim \lambda^2/\Delta \lambda$) is consistent with the used 2 nm FWHM (full width at half maximum) interference filters.} 
\end{center}
\end{figure}

However, to observe a yet more dramatic HOM coincidence dip we performed a similar measurement on a directional coupler with a nominal reflectivity of 1/2. The resulting HOM dip using coupler 3 is shown in figure~\ref{hom}. The directional coupler in question could not be directly accessed in our setup though. In order to obtain this data we sent photons to inputs $V_B$ and $C_1$.  In such a measurement some of the photons were reflected from the 1/3 reflectivity couplers, but the remaining transmitted photons interfered at the $\eta_3=1/2$ coupler as required. The effect of the additional 1/3 reflectivity couplers was simply to reduce the observed coincidence rate between fibres $T_0$ and $T_1$ by a factor of 4/9. Despite this the observed asymptotic classical coincidence rate was above 1000 per second. The quantum behaviour of the photons is evident in the figure; the coincidences nearly vanished when the optical path lengths were equal. The result is consistent with the typical measured bare source coincidence rate of 11000 when we take into account the approximately 65\% coupling efficiency (including input and output coupling and propagation loss) through the chip. One needs to also account for the fact that the number of observed coincidences through a 1/2 reflectivity coupler is precisely 1/2 of those obtained without any coupler (in the case of distinguishable photons). That is, the approximate coincidence rate in the distinguishable case should be $11000\times 1/2 \times 0.65^2 \times 4/9 \approx 1000$. The theoretical expectation taking into account the measured reflectivity $\eta_3=0.469\pm 0.009$ is $99.2\pm 0.4\%$ according to equation~\ref{vis}. Least squres fitting on the other hand yields a visibility of $95.8 \% \pm 1.6 \%$ as the experimental result. 
The experiment is inclusive of accidental/multipair coincidences which will account for less than $1\%$ of the coincidences. Thus the source visibility is limited to about 99 $\%$ due to statistics even for perfect 50/50 couplers. To summarise, the non-ideal reflectivity and multipair/accidental coincidences almost account for the observed visibility. The source of the remaining deviation is not confirmed. Small differences in the polarisation state (less than 1\%) of the photons are a possible source of error.

For the reconfigurable two-qubit gate we assume a notation which is compatible with earlier work on the CNOT. That is, the target qubit waveguides are assumed to be permutated \cite{shor}. The directional couplers with the reflectivity of $\eta$ are described by the unitary 
\begin{equation}
U_\eta=
\left(\begin{array}{cc} \sqrt{\eta}& i\sqrt{1-\eta} \\ i\sqrt{1-\eta}& \sqrt{\eta} \end{array} \right).
\end{equation}
Up to a global phase, the quantum photonic circuit implements probabilistically the gate 
\begin{equation}\label{U}
U(\phi)=\left[I\otimes U_{1/2}\right] C_z \exp\left(\frac{i\phi\sigma_z}{2}\right)\left[I\otimes (U_{1/2} \sigma_x)\right],
\end{equation}
where $C_z$ is the controlled probabilistic phase gate
\begin{equation}
C_z=|0\rangle \langle 0|\otimes I+ |1\rangle \langle 1|\otimes \sigma_z.
\end{equation}
One way to obtain this result is to consider the overall unitary $U$ describing the 6 by 6 interferometer and then project the two-particle product $U\otimes U$ to the symmetrised part of the Hilbert space corresponding to indistinguishable photons. The probabilistic gate is then a block in the resulting unitary within the logical subspace with prefactor of 1/3.  

\begin{figure*}
\begin{center}
\includegraphics[width=0.99\textwidth]{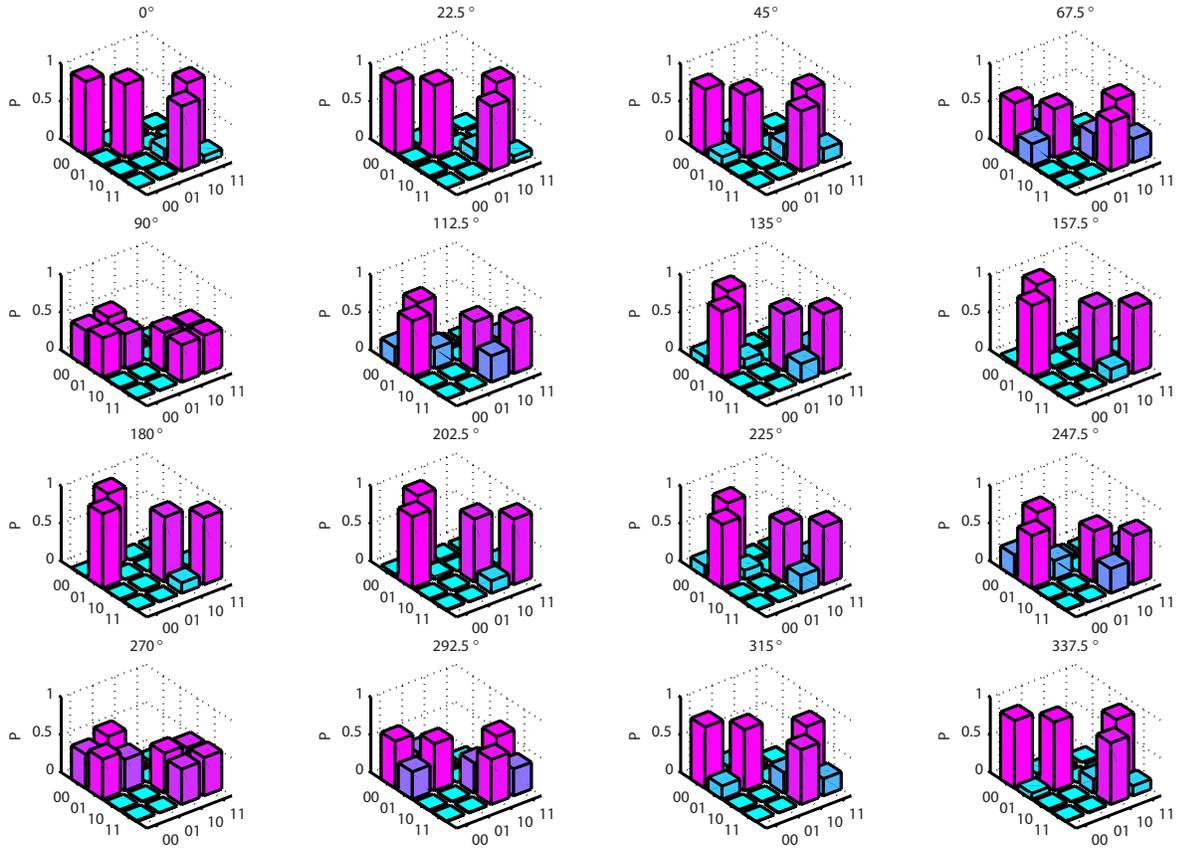}
\caption{\label{meas}Measurement of the reconfigurable two-qubit gate. The 16 panels correspond to different values of the phase shift $\phi$.  Each panel illustrates the probability of output states $|00\rangle$, $|01\rangle$, $|10\rangle$ and 
$|11\rangle$ (axis on the left side) as a function of the four input states.} 
\end{center}
\end{figure*}

\begin{figure*}
\begin{center}
\includegraphics[width=1\textwidth]{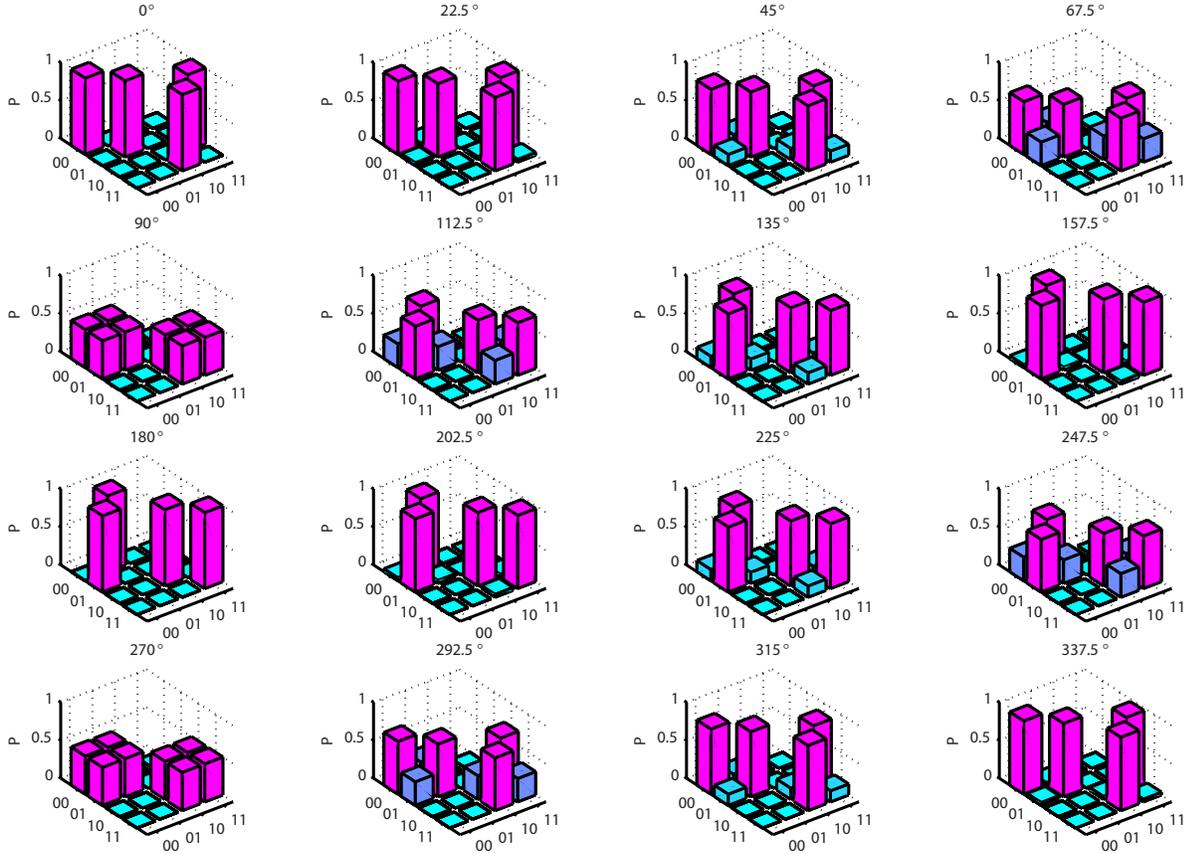}
\caption{\label{theory}Theoretical prediction for a perfectly working device with ideal reflectivities. The shown probabilities are calculated at different values of the phase $\phi$. The same result can be obtained directly form equation~\ref{U} or from general description of the $6 \times 6$ interferometer with properly symmetrised input and output states.} 
\end{center}
\end{figure*}

The reconfigurability is implemented via a rotation about the $\sigma_z$ axis of the Bloch sphere.
In the case where $\phi=0$, the circuit is similar to that in Refs.~\cite{politi,laingAPL} and we get a gate similar to the CNOT
\begin{equation}
U(0)=\left(\begin{array}{cccc} i & 0 &0 &0\\0&i&0&0\\0&0&0&1 \\0&0&-1&0 \end{array} \right).
\end{equation}
However, in the present setup the two-qubit matrix can be reconfigured to yield much more complicated forms of entangling operations. In general, the ideal theoretical expectation is 
\begin{equation}\label{mat}
U(\phi)=\left(
\begin{array}{cccc}
i\cos\left(\frac{\phi}{2}\right) &i\sin\left(\frac{\phi}{2}\right)&0 &0\\
-i\sin\left(\frac{\phi}{2}\right)&i\cos\left(\frac{\phi}{2}\right)&0&0\\
0&0&-\sin\left(\frac{\phi}{2}\right)&\cos\left(\frac{\phi}{2}\right) \\
0&0&-\cos\left(\frac{\phi}{2}\right)&-\sin\left(\frac{\phi}{2}\right) \end{array} \right).
\end{equation}

To characterise the functioning of the two-qubit gate we performed four sets of measurements corresponding to the logical input states 
$|00\rangle$, $|01\rangle$, $|10\rangle$ and 
$|11\rangle$. The optical path lengths were calibrated carefully with the help of further Hong-Ou-Mandel-type measurements when required. Such calibration needed to be done simply due to the fact that the coupling to the chip was implemented using V-groove arrays with fibres of varying length. Out of the possible coincidences between APD clicks only the events corresponding to allowed qubit logical states were kept for the logic gate analysis. The gate is considered to work whenever the two photons remain within the subspace such that there is precisely one photon in the target qubit waveguides and the control qubit waveguides. The probabilistic nature of the gate is evident in the fact that an allowed coincidence is detected with the probability 1/9. Each measurement contains 10 s worth of data.

Figures~\ref{meas} and~\ref{theory} illustrate the main result of the paper. As can be seen from the theoretical and measured probability distributions, varying the phase gradually transforms the unitary two-qubit gate from a CNOT-type gate (at $0^\circ$) to a permuted CNOT at $180^\circ$ and back. The measured data closely follows the ideal theoretical prediction in figure~\ref{theory}. The theoretical prediction was generated using the design values and does not involve any fitting. More quantitatively, we can evaluate the similarity \cite{laingAPL} between the measured operations $M$ and ideal ones $I$ using the expression 
\begin{equation}
S=(\sum_{k,l}\sqrt{I_{kl}M_{kl}})^2/16.
\end{equation}
Averaged over all the 16 data sets, we get a similarity of $97.7\% \pm 2.1\%$. Such a high degree of similarity between the ideal and measured gates is very encouraging. Using the independently measured reflectivities instead of the design values for a theretical simulation yields only a slightly improved similarity of $98.1\% \pm 1.3\%$. The individual similarities range from 0.951 (at $\phi=0$) to 0.993 (at $\phi=90^\circ$). Considering the small imperfections in the HOM measurement suggests that some of the remaining error is due to imperfect quantum interference.  It is also likely that non-identical output coupling losses and detector efficiencies can cause slight errors, along with noise in the phase shifter. 

The applications of the device can be understood by considering equation~\ref{mat} and the illustrations in figures~\ref{meas}-\ref{theory}. Among other things, we are able to reconfigure the chip to implement a logic "0"-controlled operation such that performing a NOT on the target qubit is conditional on the control qubit being in the state 0 rather than 1 as usual.
These two cases correspond to $\phi=180^\circ$ and $\phi=0^\circ$, respectively. Furthermore, non-entangled states such as 
$1/\sqrt{2}(|0\rangle \pm i|1\rangle)\otimes |0\rangle$ can be transformed to generate any four of the maximally entangled Bell states \cite{bell} using the present device and the aforementioned choices of $\phi$.
Perhaps the most intriguing application of device though is the ability to partially entangle states to perform weak measurements tunably, as was done in bulk optics recently \cite{pnas}.

Our results constitute a proof-of-principle demonstration of a reconfigurable two-qubit entangling gate on a photonic chip. From the point of view of compiling quantum algorithms it is always desirable to have access to a more flexible set of two-qubit gates, instead of e.g. the standard CNOT. To this end, we have demonstrated high fidelity implementations of a variety of probabilistic two-qubit gates on a photonic chip. Our results speak strongly in favour of integrated linear optics as a platform for quantum computing. The main drawback in the qubit-based linear photonics approach is the fact that the gates have to be probabilistic. An interesting and natural future direction for photonics-based quantum information is the use of purely bosonic models for computation.

\pagebreak

\end{document}